\documentclass[aps,pre,floatfix,twocolumn,superscriptaddress]{revtex4-2}
\usepackage{amsmath}
\usepackage{amsfonts}
\usepackage{mathrsfs}
\usepackage{epsfig}
\usepackage{graphicx}
\usepackage{dcolumn}
\usepackage{hyperref}
\usepackage{xcolor}
\usepackage{placeins}
\usepackage{subfigure}
\usepackage[hyphenbreaks]{breakurl}
\hypersetup{
    colorlinks=true,
    linkcolor=black,
    filecolor=black,      
    urlcolor=black,
    pdfpagemode=FullScreen,
    citecolor=black,
    }

\begin{document}
\title{Environmental and cell-cell signaling shape developmental trajectories across morphogenetic landscapes}
\author{J Hareesh}
\affiliation{The Institute of Mathematical Sciences, CIT Campus, Taramani, Chennai 600113, India}
\affiliation{Homi Bhabha National Institute, Training School Complex, Anushaktinagar, Mumbai 400 094, India}
\author{Sitabhra Sinha}
\affiliation{The Institute of Mathematical Sciences, CIT Campus, Taramani, Chennai 600113, India}
\affiliation{Homi Bhabha National Institute, Training School Complex, Anushaktinagar, Mumbai 400 094, India}

\date{\today}
\begin{abstract}
Despite the variability in gene regulation and environmental conditions, development of an organism occurs through a sequence of highly coordinated patterning processes. Cells integrate different signals to accurately infer their position in order to adopt an appropriate identity. Using a model of epigenetic landscape originally proposed by Waddington to describe cell-fate determination, we establish the critical role played by juxtacrine signaling between cells in determining tissue patterns. Subsequently we systematically coarse-grain the model at the tissue scale to map its patterning  to transition
between states in a binary spin model having a free energy landscape. We show that such landscapes serve as a powerful unifying framework for describing development of biological systems across distinct spatio-temporal scales.
\end{abstract}
\maketitle
\section{Introduction}
Development in living systems comprises a series of highly coordinated patterning processes, whose dynamics are governed by interactions between its components and its environment~\cite{Wolpert2015, Turing1952, Thompson1942, Kondo2022, Oster1988, Zhang2018, Gilbert2018, Maroto2012, Meinhardt1982, Pathak2020, Schnell2002}. At the cellular scale, the relevant information is often interpreted and translated into appropriate actions through signaling pathways
that are sensitive to chemical concentrations. The input may take the form of morphogen molecular gradients or bound receptor-ligand complexes enabling contact-mediated interactions with neighboring cells~\cite{Wolpert1969, KERSZBERG2007205, Gurdon2001, Zhu2004, annurev:/content/journals/10.1146/annurev.cellbio.18.012502.083458, Wigglesworth1940, Aulehla2010, Gierer1972, Koch1994, Meinhardt2000}. Despite the variability in these systems and the environment that they operate in, developmental trajectories and the phenotypic outcomes are surprisingly precise~\cite{Arias2006, Kitano2004, Lagha2012}. This is vital for survival of the organism as minor errors in the early developmental stages can get amplified over successive stages and lead to non-viable phenotypes \cite{MEINHARDT1983375, Mugler2016}. Even in cases where they are viable, pathological variations may lead to hampered functionality and lowered fitness for individuals. Much of decision-making during development, and the ability to canalize variability in the interactions, reflect the ability of the organism to appropriately switch between distinct phenotypic outcomes separated by boundaries that
are akin to critical transition lines demarcating distinct phases in physical systems. There is increasing evidence that many biological systems are indeed poised near critical points, making drastically diverse responses accessible to the system~\cite{doi:10.1073/pnas.1324186111, mora2011biological, bak1997nature, PhysRevLett.71.4083, 10.1098/rsta.2007.2092, beggs2003neuronal, doi:10.1073/pnas.97.7.3183, doi:10.1073/pnas.1005766107, PhysRevLett.90.158101}. Although there are some obvious benefits in being fine-tuned near criticality, such as enhanced sensitivity to external signals, understanding the mechanisms that enable biological systems to be driven to criticality in reasonable evolutionary timescales, as well as, to retain criticality against mutations or fluctuations in the environment remains a significant challenge~\cite{doi:10.1073/pnas.95.15.8420, Stanoev2020}. 

To address these questions, we focus here on a low-dimensional representation of cellular dynamics rather than on detailed microscopic models of its specific aspects. Such an approach has led to many theoretical insights and practical applications in different domains of biology~\cite{Senior2020, HALABI2009774, rouviere2024unifiedviewallostery, FERRELL2012R458, Wang02012015, 10.1063/1.3478547, doi:10.1073/pnas.0910331107, doi:10.1126/science.1749933, doi:10.1126/science.7886447, wang2006funneled, Wu2017, Brown2003}. Beyond the convenience of being tractable, low-dimensional models are also more meaningful from the perspective of the biological system we intend to analyze~\cite{10.1098/rsfs.2022.0002, stanoev2021robustness}. In the
absence of canalization over certain parameters (or their combinations) upto varying degrees, it is not possible for cells to generate robust responses in an otherwise extremely high-dimensional space~~\cite{levins2009}. The concept of \textit{epigenetic landscapes} originally proposed
by Waddington [Fig.~\ref{fig1}~(a)] serves as an excellent framework to coarse-grain the complicated dynamics of a cell to an appropriate reduced dimensional space without losing essential insights into the causal mechanisms involved in cell-fate determination~\cite{Waddington1957, rand2021geometry, corson2017gene, doi:10.1073/pnas.1408561111, doi:10.1073/pnas.1017017108}. We note that free energy functions are an appealing analogy to model these landscapes, even though they are
out of equilibrium: constantly evolving, modulated by its own internal state and driven via coupling with environment~\cite{Cross1993}. By stepping away from a detailed consideration of cellular energetics, we can derive a dynamical system description that can be mapped to a much simpler, analogous  equilibrium phenomenological model. Thus, free energy functions, being amenable to techniques from statistical physics, enable us to formalize a minimal description of cell fate determination in the presence of morphogens (global field) and juxtacrine signaling (interactions with neighbors).

Following the French Flag paradigm~\cite{Sharpe2019}, we consider the case of tissue segmentation in the presence of a morphogen gradient~\cite{Dahmann2011, Gurdon2001, Ashe2006, Dessaud2008, Driever1988, Lander2009, Lewis1977, Meinhardt2009, Tickle1999, Vincent2001}. Undifferentiated cells adopt any one of two distinct identities $A$ or $B$, depending on their position in the embryo that is inferred from the local morphogen concentration~\cite{Zagorski2017, Dubuis2013, Lander2013}. This results in the formation of two domains having some desired proportions, beginning from an otherwise homogeneous tissue. In this paper, we use the free energy of a system of binary spins arranged along
an array to model the epigenetic landscape, with the spin states 
corresponding to cell fates, the morphogen gradient to an external space-varying magnetic field and juxtacrine signaling between cells to ferromagnetic interactions~\cite{Kuyyamudi2022,Muhamet2025}. In the absence of cell-cell interactions, we show that a cell needs to be fine tuned near criticality for robustness, which comes at the expense of extended developmental time. Both of these features can severely compromise the fitness of the organism. Introducing receptor-ligand interactions between cells enable the organism to overcome these problems. Surprisingly, we also find that selection pressure for increased precision and reduced patterning time forces the system to self-organize to the critical regime. Our results show that free energy landscapes resembling those suggested by Waddington
can serve as a powerful unifying framework in biology to study development and evolution, as it
allows us to map distinct, unrelated biological mechanisms to very similar operational principles
We also show how to obtain a much simpler description for patterning at the tissue scale by systematically coarse-graining cellular scale descriptions of the process.
\section{Energy landscape: Cellular Scale}
\begin{figure}[tbp]
\includegraphics[width = \columnwidth]{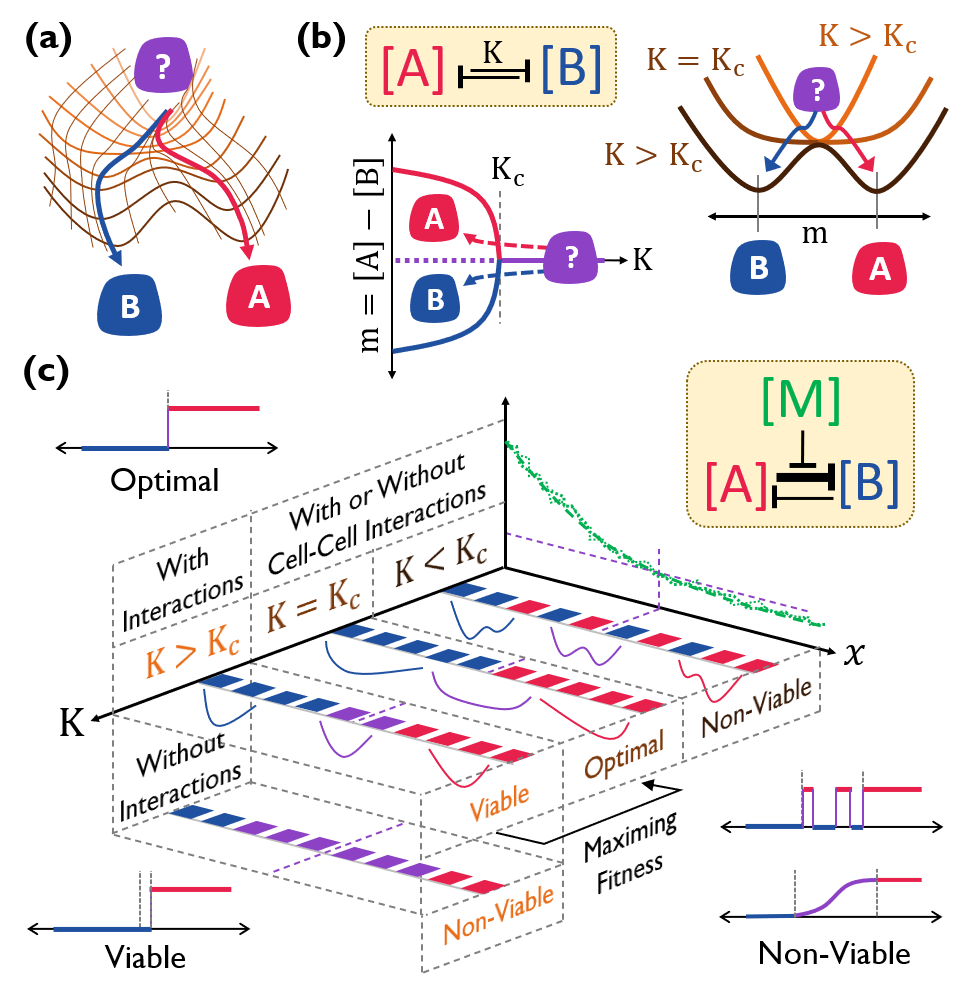}
\caption{Waddington landscape as a conceptual framework to describe cell fate determination during development. (a)~Dynamics of fate choice represented as trajectories in the epigenetic landscape. An undifferentiated cell (violet) differentiates into one of the two possible cell types $A$ (red) or $B$ (blue). (b)~A toggle switch comprising two representative genes that correspond to the two distinct cellular identities. Inhibition threshold $K$ determines the genotypic outcomes. Co-expression regime $(K>K_c)$ is interpreted as the undifferentiated cell state while the bi-stable regime $(K<K_c)$ corresponds to differentiated cell states. A quartic potential effectively models the landscape. 
(c)~Spatial gradient of a morphogen cues the cell to preferentially adopt distinct fates in accordance with their spatial position, leading to tissue segmentation. The morphogen modifies the potential function to bias it either towards $A$ or $B$. Optimal segmentation refers to the emergence of domains of equal proportions. The pattern is said to be viable if it has two domains, but the interface is displaced by at most a single cell. Patterns having more skewed proportions or jagged/ multiple interfaces are considered to be non-viable. In the absence of cell-cell coupling, optimal patterns are obtained only at criticality $(K \sim K_c)$, while in the presence of cell-cell coupling, a viable phenotype is seen over a broad range of parameters $(K \geq K_c)$. Selection for increased precision can tune the system, driving it from a viable regime $(K \geq K_c)$ to criticality $(K \rightarrow K_c)$.}
\label{fig1}
\end{figure}
In the course of development, cells commit to distinct fates by inferring their position, which usually requires interpreting chemical concentrations from morphogen gradients~\cite{DRIEVER198895, Yuste2010, Kicheva2007, Lander2002, Lander2007, Saka2007, Wartlick2009}. We consider a $1$-dimensional array of cells, our results can be easily generalized to higher dimensions. The 
normalized morphogen concentration is modeled as an exponentially decaying function of position $l$ in a domain of size $L=1$, viz., $M(l) = \exp(-l/\lambda_M)$~\cite{Crick1970}. We consider each
cell to have two mutually inhibiting patterning genes, $A$ and $B$. In the absence of morphogen$(l \approx 1)$, $A$ dominates [Fig.~\ref{fig1}~(b)]. On the other hand, morphogen concentration above a threshold $K_m$ amplifies the inhibition of $A$ resulting in up-regulation of $B$ $(l \approx 0)$:
\begin{equation}
    \Dot{A} = \alpha \Theta_h(K_1,B) - A/\tau +A\eta^t,
\end{equation}
\begin{equation}
    \Dot{B} = \alpha \Theta_h(K_2,A) - B/\tau +B\eta^t,
\end{equation}
where $\Theta_h(a,b) = a^h/(a^h +b^h)$ and the gating variable $h = 2$. We use normalized gene expression values $(\alpha \tau = 1)$, assuming identical production rates $\alpha$ and degradation rates $\tau$ for the genes. Further, we choose $K_1 = K$, $K_2 = 2K\Theta_h(M,K_m)$ and $K_m = e^{-1/2\lambda_m}$, to conveniently position the interface at the midpoint of the domain, $l = 1/2$. The multiplicative noise factor $\eta^t$ follows Brownian statistics $<\eta^t>=0$, $<\eta^t \eta^{t+\Delta t}>=2\sigma^2\delta(\Delta t)$.

The system can operate in a bistable regime, where the expression of one gene dominates the other, or a co-existence regime where both genes are expressed at comparable levels. We infer these two regimes as differentiated and undifferentiated regimes respectively. The system can be fine tuned to operate at a critical regime, where noise can induce strong temporal fluctuations in the gene expression, but on average, both the genes are expressed equally [Fig.~\ref{fig1}~(c)]. Near criticality, after discarding transient behavior, the dynamics of the genes is almost confined along the slow eigen-direction and we can study the dynamics of an order parameter $m = A-B$ without any significant loss of information~\cite{10.1063/1.4923066}. As the dynamics is confined to a particular eigen-direction, the gene expression of $A$ and $B$ are anti-correlated, such that, $A+B=1$. Although, the analysis that follows is valid near the critical point, the qualitative features of the dynamics are preserved even away from criticality. At the interface, assuming $K \leq K_c$ $(K_c=1/2)$, the difference in gene expression at steady state is $m_* = \pm 2\sqrt{K_c^2-K^2}$. The steady state distribution can be expressed as $P(m) = \delta\left(m^2-m_*^2\right)/2$. We can approximate the $\delta$-function with a Gaussian distribution whose variance can be interpreted as the gene expression noise which is tolerated in a valid cell-type identity. If we assume that there exists a free energy function $\phi(m)$ for which $P(m) = \exp\left({-(m^2-m_*^2)^2/(2\sigma^2)}\right)/(2\sigma\sqrt{2\pi})$ is the equilibrium description, $P(m) \sim e^{-\beta \phi(m)} \implies \beta \phi(m) \sim -\ln{P(m)}$. Ignoring constant terms, we obtain,
\begin{equation}
    \phi(m) = \frac{m^4}{2\beta\sigma^2} - \frac{4(K_c^2-K^2)m^2}{\beta\sigma^2},
\end{equation}
where $\beta \sim 1/T$ is the inverse temperature which is related to the fluctuations in the underlying signaling networks. It is important to note that this effective temperature (even though it is related to
) is not the same as the physical temperature that molecules experience within the cell.

When the effect of the morphogen gradient is taken into account, the energy function has to be augmented with a spatial gradient field term. As we are interested in the robustness of cell fate determination near the segmentation boundary, we will work with a first order expansion that is valid near the interface. We further assume first-order perturbations near criticality to capture the qualitative behavior of both bistable and co-expression regime:
\begin{equation}
    \label{cellenergyfunc}
    \phi(m) = \frac{m^4}{2\beta\sigma^2} - \frac{4(K_c^2-K^2)m^2}{\beta\sigma^2} - \frac{7mx}{8\beta \lambda_m \left(K_c-K\right)^2},
\end{equation}
where the new spatial coordinate $x = l-1/2$ is chosen to conveniently place the interface at the origin. The quartic potential form in Eqn.~\ref{cellenergyfunc} indicates bistability, while the linear term in $x$ captures the effect of the morphogen field. This is reminiscent of the free energy function of the Ising Model, where the quartic term leads to symmetry breaking at low temperature and the linear term is the bias due to an external field. The contribution of morphogen is weak for larger wavelengths $\lambda_m$. However, the weak signal can be amplified near criticality $K\rightarrow K_c$. If we assume that the morphogen gradient scales with the tissue size, the interface robustness is significantly compromised for larger domains~\cite{Morton2010}. Hence, mechanisms of robustness also constraints the permissible diversity in the size of organs and organisms in a population.

Following the Ising model analogy, we replace a cell with a set of $N$ interacting spins [Fig.~\ref{fig2}~(a)], where each spin can take any one of two states $\pm 1$, and the net magnetization $(m = \sum_i \sigma_i/N)$ labels distinct cell states. The energy of the spin system is given by
\begin{equation}
    E = -\frac{J}{2N}\sum_{ij}\sigma_i \sigma_j - \sum_i h(x) \sigma_i,
\end{equation}
where $x$ is to the spatial position of the cell which has been replaced by the spin system. 
The system exhibits bistability at low temperatures (or with strong interactions), where the magnetization spontaneously goes to one of the two absorbing states $m \approx \pm1$. At
higher temperatures (or for weaker interactions), it remains in a non-magnetized state $m \approx 0$   
[Fig.~\ref{fig2}~(a), right]. The former is interpreted as a differentiated cell, while the latter as an undifferentiated cell. At the critical temperature, the system exhibits extremely slow dynamics with large fluctuations around $m = 0$. Although this is a phenomenological description of two mutually inhibiting genes, the system is equally representative of any interaction network capable of storing distinct genotypes. The susceptibility $\chi = \partial m/\partial h$ is a response function that characterizes the change in magnetization due to small variation in the external field and 
diverges near the critical point $(T \rightarrow T_c  \implies\chi \rightarrow \infty)$ implying extreme sensitivity to external signals.

\section{Dynamics in the absence of inter-cellular signaling}
\begin{figure}[tbp]
\includegraphics[width = \columnwidth]{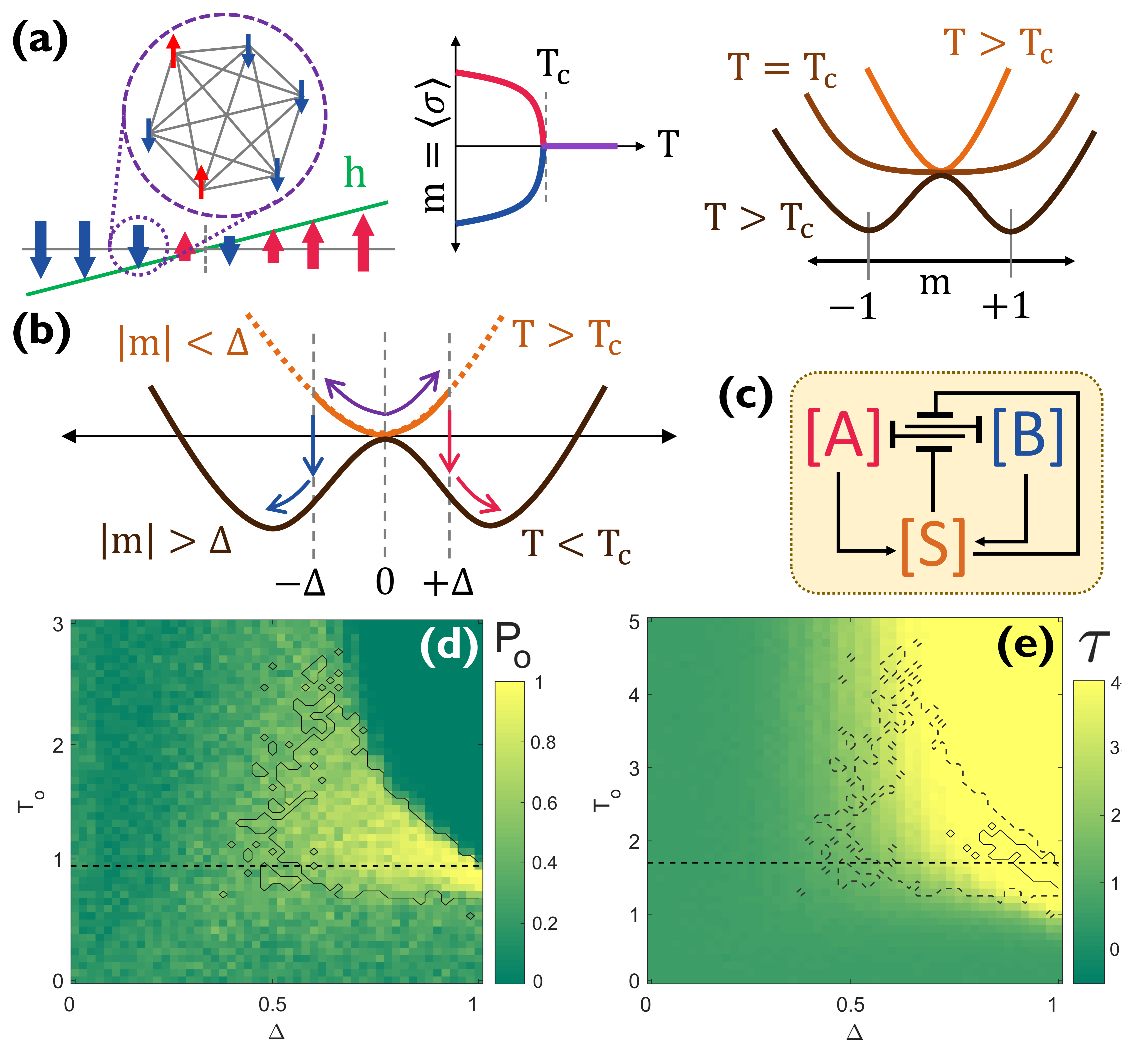}
\caption{In the absence of any inter-cellular signaling, the decision to switch from an undifferentiated to a differentiated state is determined by the internal state of the cell and the external morphogen concentration. (a)~A cell is modeled as a set of binary spins $(\pm1)$ with ferromagnetic interactions, where the average magnetization corresponds to the cell-state. The effect of the morphogen appears as a spatial gradient in the external magnetic field $h$ (green). High temperature $T>T_c$ corresponds to an undifferentiated state, while low temperatures $T<T_c$ corresponds to a differentiated state. Free energy of the Ising model effectively describes the epigenetic landscape. 
(b)~Decision to commit to a fate depends on how primed the cell is towards a particular fate. Once the magnetization is pushed past a threshold $\Delta$ due to external field or fluctuations, reducing the temperature from $T_o$ to $0$ remodels the landscape. (c)~A representative gene regulatory motif that modulates the landscape geometry by tuning the mutual inhibition between $A$ and $B$ via an enzyme $S$. Dominant expression of one gene over the other leads to reduced concentration of $S$, which strengthens the inhibition between the genes, resulting in the cell fate being fixed. 
(d-e)~Probability $P(\delta=0)$ of obtaining an optimal pattern (e) and the time $\tau$ (in
logarithmic scale)
required for patterning to be achieved (e) shown as function of $\Delta$ and $T_o$ of the undifferentiated cell state. The contours indicate the region where $P(\delta\leq1)>0.9$. Robust patterning is observed in the limit $\Delta \rightarrow 1, T_o \rightarrow T_c$. 
Robustness is seen to occur at the expense of extending the patterning time. Results shown
are averaged over $50$ realizations, with each being evolved for $10^4$ MC time-steps.}
\label{fig2}
\end{figure}
We consider a scenario where the cells have to commit to one of the two distinct fates solely based on their own internal state and the external morphogen signal. We assume a regulatory scheme that modifies the underlying epigenetic landscape once the cell is primed towards a certain 
fate~\cite{waddington1940organisers}. Once committed, the fate is retained by switching from the co-existence regime to the bi-stable regime [Fig.~\ref{fig2}~(b)]. Modulating the catalysis of chemical reactions that regulates the strength of mutual inhibition between the patterning genes would allow such modifications of the energy landscape. For instance, when the patterning genes are co-expressed, it enhances the production of an effector molecule. It inhibits the enzymes that catalyze mutual inhibition, thereby reducing the inhibition strength to a basal value. The cell commits to a particular fate because of noise in the gene expression and the bias induced by the morphogen signal~\cite{PhysRevLett.88.048101}. This leads to depletion of effector molecules, resulting in reduced inhibition of the enzyme that facilitates mutual inhibition between $A$ and $B$, allowing bistability to retain the cell type it was primed for [Fig.~\ref{fig2}~(c)]. Evolution of the effector molecule concentration $S$ is given by
\begin{equation}
    \Dot{S} = \alpha_2 \Theta_g(A,K_o)\Theta_g(B,K_o) - S/\tau_2.
\end{equation}
Assuming fast response $(\tau_2 \ll \tau)$, the effector molecule concentration follows the dynamics of $A$ and $B$, viz., $S_* = \alpha_2 \tau_2 \Theta_g(A,K_o) \Theta_g(B,K_o)$. The inhibition threshold $K$ for patterning genes is given by $K_{1,2} = (K_o - K_l)S/(\beta \tau_2) + K_l$, where $K_l \ll K_c$ corresponds to the cell operating at the bistable regime, $\alpha_2$ is the production rate, and $g$ is the gating variable. As the effector molecules modulate the inhibition strength, the analogous response function for a spin system would be the dependence of interaction strength or temperature on the magnetization of the system. At weak magnetization, $|m|<\Delta$, the interaction strength reduces, or equivalently, the temperature increases. The temperature of the spin system can therefore, be expressed as a function of the magnetization, $T = T_o\Theta_{g}(\Delta, |m|)$.

We quantify the patterning error $\delta$ as the width of the region that deviates from the optimal pattern. We examine the probability of the patterns being optimal $P_o = P(\delta=0)$ and viable $P_v = P(\delta=1)$ by varying thresholds and temperature that describes the undifferentiated cell state. At low temperatures, the system is operating in a bistable regime. This prevents convergence to the desired state in reasonable time-scales and the bias in initial conditions dominate the final outcome. At higher temperatures, the fluctuations are not strong enough to cross the threshold $\Delta$, and the system fails to commit to a fate $(m\sim0)$. On the other hand, if the threshold is low, the system can converge to undesired states due to spurious fluctuations and bias introduced by arbitrary initial conditions. However, at the critical temperature, there is robust convergence because of high susceptibility and large fluctuations that facilitates crossing arbitrarily high thresholds. In fact, at criticality, the higher the threshold, the less likely it is to cross the threshold against the field because of diverging susceptibility. This mechanisms allows the cell to amplify the effect of weak external field but only in the limit $\Delta \rightarrow 1, T_o \rightarrow T_c$, implying that for robust patterning, the underlying regulatory network must be fine tuned near criticality and large response thresholds [Fig.~\ref{fig2}~(d)]. 

In the \textit{Drosophila} embryo, gap-gene expressions near the segment interface along the anterior-posterior axis exhibits several features that are characteristic of criticality, such as strong anti-correlation in the gene expression profile~\cite{doi:10.1073/pnas.1324186111}. However the parameter regimes away from criticality where the patterns are viable are quite restricted and near-critical systems cannot evolve through gradual increments in fitness if they begin far from criticality. 
The explanation for the existence of near-critical systems would need to rely on random chance mutations. Even if that were the case, the cumulative effect of random mutations or environmental shocks over evolutionary timescales could perturb the system away from criticality, making it difficult to retain the fine-tuning which is essential for robustness. More importantly, if cells rely on internal mechanisms to modulate their states, robustness comes at the expense of developmental time, which is detrimental to the fitness of the individual [Fig.~\ref{fig2}~(e)]. Hence we need to turn to
other mechanisms, specifically, signaling between cells.

\section{Energy landscape: Tissue Scale}
Robustness of patterns can be enhanced by integrating spatial information via signaling across 
cells~\cite{PhysRevE.103.062409, 10.1242/dev.197608, sprinzak2010cis, monk1997cell, nichols2009suppression, saiz2020growth, collier1996pattern, 10.1242/dev.199926, PhysRevE.106.L022401, PhysRevE.106.L022401, Whited2019, Barad2010, Basch2016, Boareto2015, Caneparo2011, Carmona2008, Classen2005, Cohen2010, Ellison2016, Erdmann2009, Heitzler1991, Roy2015}. We utilize the Eph-Ephrin pathway as the means for contact mediated cell-cell communication in our model~\cite{10.1242/dev.038935, Kullander2002, Davy2005}. The pathway is ubiquitous to several signaling processes in a wide range of systems and is known to play a role in tissue segmentation~\cite{COOKE2005536, COOKE2002260}, viz., perturbing the pathway leads to improper boundary formations or even the absence of boundaries altogether~\cite{Mellitzer1999}. We consider a gene regulatory motif where the receptor $R$ (ligand $L$) gets uniquely expressed in cell type $A$ $(B)$. Binding between receptor and ligand leads to bi-directional signaling, thereby upregulating effector molecules $S$ in cells at the interface~\cite{Holland1996, KLEIN1999R691}. The effector molecules reduce the inhibition between the patterning genes [Fig.~\ref{fig3}~(a)], remodeling the landscape to an undifferentiated state~\cite{Wilkinson04072014, Stewen2024}. The dynamics of receptor, ligand and effector molecules are given by
\begin{equation}
    \Dot{R} = \alpha_2\Theta_h(A,K_o)\Theta_h(K_o,B)-k_{tr}RL_{tr} - R/\tau_2,
\end{equation}
\begin{equation}
    \Dot{L} = \alpha_2\Theta_h(B,K_o)\Theta_h(K_o,A)-k_{tr}R_{tr}L - L/\tau_2,
\end{equation}
\begin{equation}
    \Dot{S} = \alpha_3 k_{tr}\left(RL_{tr}+R_{tr}L\right)/2 - S/\tau_2,
\end{equation}
where $K_1 = \left(\left(2K-K_l\right)S+K_l\right)\Theta_h(M,K_m)$, $K_2 = \left(\left(K-K_l\right)S+K_l\right)$ with $K_l\ll K_c$,
$k_{tr}$ is rate of receptor-ligand binding across neighboring cells, and $\tau_2$ is the degradation rate. We assume fast dynamics for the receptor-ligand pathway. $(\tau_2 \ll \tau)$.

We consider a pair of cells with state variables $m$ and $m_{tr}$. When the cell types are different, $(m_{tr}m<0)$, the signaling modifies the landscape to an undifferentiated state, driving the system to $(m, m_{tr}) \rightarrow (0,0)$. When they are the same type $(m_{tr}m>0)$, the signal is absent and the cell fates are retained [Fig.~\ref{fig3}~(b)]. The steady state solution is given by $(m,m_{tr}) =\pm(1,1)$ [Fig.~\ref{fig4}~(a)]. The effect of juxtacrine signaling can be taken into account with an additional ferromagnetic interaction term in the free energy of the cells, $\phi(m) \rightarrow \phi(m) - Jm_{tr}m$.
\section{The effect of Juxtacrine Signaling}
\begin{figure}[tbp]
\includegraphics[width = \columnwidth]{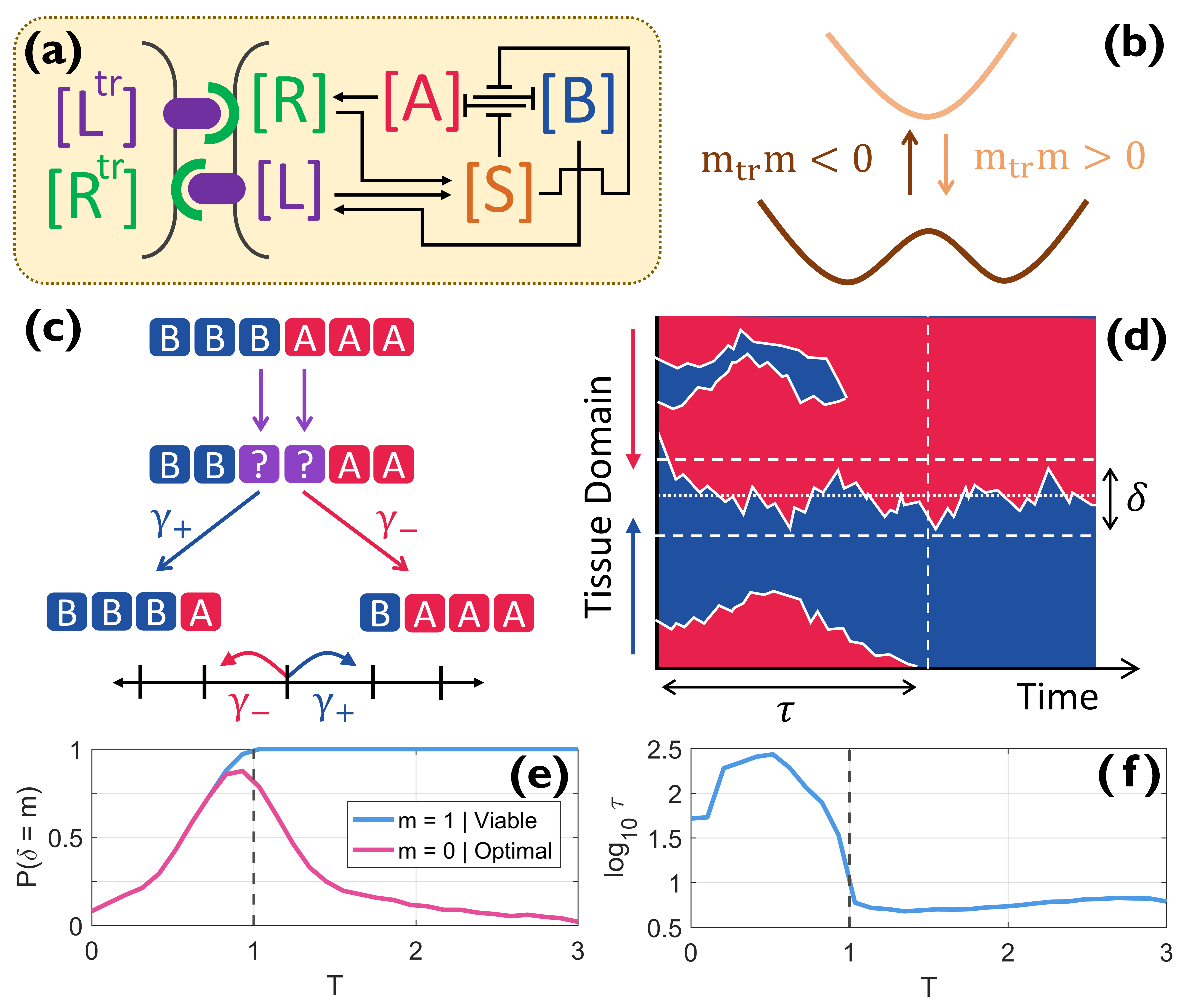}
\caption{In the presence of cell-cell coupling, the landscape integrates signals from the morphogen, as well as, the cellular neighborhood. 
(a)~A representative gene regulatory motif motivated by the Eph-Ephrin signaling pathway, that modulates the landscape geometry. Cells of type $A$ and $B$ expresses receptor $R$ and ligand $L$ respectively. Cells that differ from their neighboring cell types have upregulated effector molecule concentration $S$ because of bidirectional signaling from bound receptor-ligand complex. (b)~The spin analog for the proposed juxtacrine signaling pathway involves decreasing the temperature to $0$ when surrounded by similar cell types $m_{tr}m>0$ and increasing the temperature to $T_o$ when surrounded by differing cell types $m_{tr}m<0$. (c)~Cells at the interface of distinct domains revert to an undifferentiated state, allowing them to switch identity. Effectively, this results in the interface being shifted to the right or left with rates $\gamma_{\pm}$. (d)~The net drift $\gamma_+ - \gamma_-$ is positive for $x<0$ and negative for $x>0$. This localizes the interface at the position where $h \sim 0$. The interfaces can either meet and annihilate or be absorbed in the domain boundaries until there is only a single interface. (e)~At low temperatures, the rates $\gamma_{\pm}$ are very low and the relaxation time diverges. Errors are determined largely by the initial conditions. At critical temperature, the interface localization is very precise, leading to optimal patterns. At high temperatures, the interface loses its precision but the patterns still remain viable. (f)~Patterning time reduces with increasing temperature due to increased mobility of the interface.}
\label{fig3}
\end{figure}
Essentially, the inhibition for cells near domain interfaces are relaxed as a consequence of effector molecules being produced through cell-cell signaling [Fig.~\ref{fig3}~(c)]. This motivates a different choice for temperature that depends on the cell states $m$ and the neighboring cell state $m_{tr}$, $T = T_o\Theta_{g}(-m_{tr}.m)$ where we follow the convention $\Theta_g(x) = 1/(1+e^{-gx})$, and the subscript $tr$ refers to neighboring cells. Let us consider the case $g \gg 1$. Away from the interface, the cells do not change their state. At the interface, cells may or may not change their state depending on the temperature $T_o$. At low temperature $T_o < T_c$, the system cannot cross the barrier of the quartic potential and remains frozen. Thus, the interface has low mobility and the errors induced by the bias in initial conditions are retained. Near criticality, $T \sim T_c$, the system is extremely precise because of large susceptibility and optimal interface mobility. At high temperature $T_o > T_c$, the interface has high mobility but the precision of the interface is compromised due to reduced sensitivity to the weak field [Fig.~\ref{fig3}~(d)]. The patterns are still viable as there are two neat domains of slightly variable proportions, with the interface remaining sharp [Fig.~\ref{fig3}~(e)]. Therefore, mechanisms that integrate spatial information need not rely on fine tuning of the dynamical parameters, as there is a broad range of parameters $(T\geq T_c)$ where the interface mobility results in a viable phenotype.

We can write down the dynamics of the interface by estimating the probability of the cells near the interface to switch their fates:
\begin{equation}
    \label{rate_exp}
    \gamma_{\pm}(x) = \gamma_o \exp{\left(\mp\sqrt{3}\sqrt{\beta (\beta -1)}hx + \frac{21}{16}\frac{\beta^2}{(\beta -1)} h^2 x^2\right)}.
\end{equation}
The transition rates indicates the presence of an effective restoring force centered at the origin, $(x>0 \implies \gamma_->\gamma_+)$ and $(x<0 \implies \gamma_+>\gamma_-)$. The convergence time is given by the inverse of the scaling factor, $\gamma_o$: 
\begin{equation}
    \gamma_o = \exp{\left(-\frac{3}{4}\left(\frac{\beta-1}{\beta}\right)^2\right)}.
\end{equation}
At low temperature limit, $T \rightarrow 0$, $\beta \rightarrow \infty \implies \tau \sim \gamma_o^{-1} \rightarrow \infty$. Not surprisingly, the convergence is also fast at the boundaries and slow for $x \sim 0$. As the temperature increases, the cell fate determination time reduces in contrast to the case of cells not interacting with their neighbors [Fig.~\ref{fig3}~(f)].

The steady state distribution near the critical temperature is given by
\begin{equation}
    p(x) \sim \frac{2 \cosh(\beta' hx)^{-1/\beta'h}}{\gamma_o\left[\exp\left(\beta'hx +qh^2 x^2\right)+\exp\left(-\beta'hx +q h^2x^2\right)\right]},
\end{equation}
where $\beta' = \sqrt{3\beta(\beta-1)}$ and $q = 21\beta^2/16(\beta -1)$. Near the critical temperature, the variance tends to zero $(\beta \rightarrow \beta_c = 1 \implies \beta' \rightarrow 0 \implies p(x) \rightarrow \delta(x) \implies)$, indicating extreme precision in the localization of the interface. 
\section{Mapping tissue patterning to spin system}
\label{section_final_ising}
\begin{figure}[tbp]
\includegraphics[width = \columnwidth]{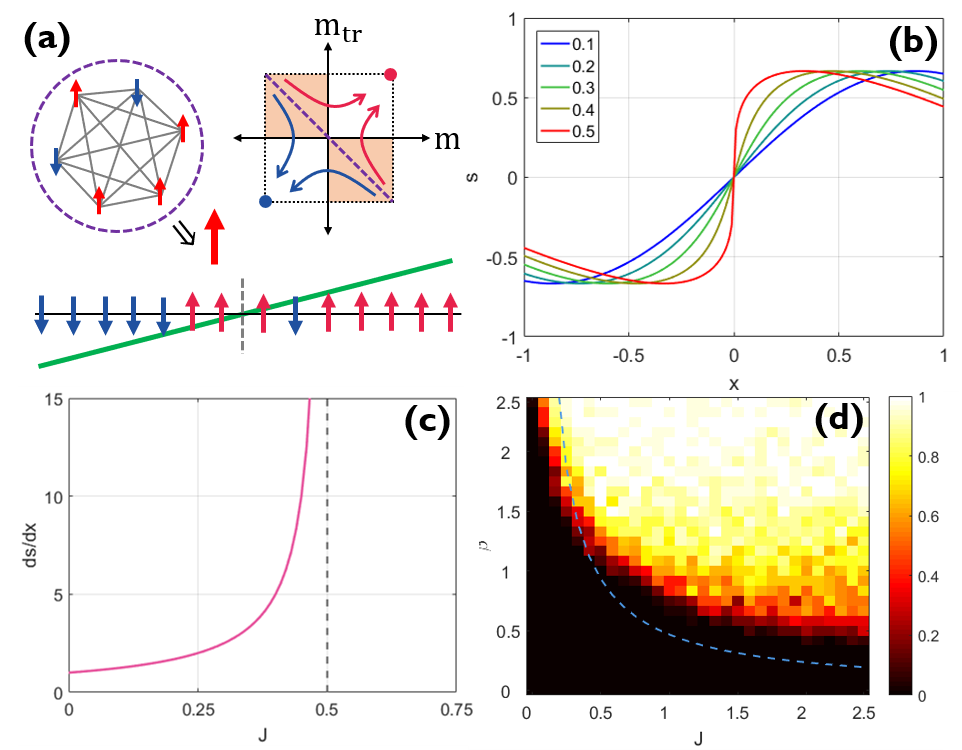}
\caption{Coarse-graining the model from cellular scale to tissue scale. (a)~Each cell is replaced by
a cluster of interacting spins that can be viewed as a ``meta-spin''. This reduces the model to a $1$-dimensional Ising model with an external field gradient. In the shaded regions $m_{tr}m<0$, the temperature $T_o \geq T_c$ resulting in $(m, m_{tr}) \rightarrow (0,0)$, while in the other two quadrants $(m_{tr}m>0)$, the system converges to $\pm(1,1)$ due to low temperature. The contact mediated interactions can be reduced to an effective ferromagnetic interaction between the meta-spins. (b-c)~Mean-field solution predicts that the interface width will reduce as the interaction strength increases, and the slope of the interface diverges as $J \rightarrow 1/(2\beta)$. (d)~Probability of obtaining viable patterns $P(\delta\leq1)$ as the inverse temperature $\beta$ and interaction strength $J$ is varied. The broke curve indicates the mean field solution that demarcates the region in 
$J-\sigma$ parameter space where optimal patterning is expected.}
\label{fig4}
\end{figure}
Trans-regulatory interactions are present only when the cell is in contact with another cell of a different type. Hence, the interaction strength between the $i^{th}$ and $j^{th}$ cells should depend on the cell state $S_i, S_j$ yielding $J_{ij} = J(1-S_i S_j)$. At low temperatures, we can coarse-grain each cell as a cluster of interacting spins (meta-spin) owning to its bistability $(S_i = \Theta_\infty (m_i) = \pm1)$. Substituting in the Ising model Hamiltonian and ignoring the constant terms, the effective energy function has exactly the same form as the $1$-dimensional Ising model in the presence of external field, viz.,
\begin{equation}
    E = -J\sum_{ij}\sigma_i\sigma_j - \sum_i h_i \sigma_i.
\end{equation}
Using Boltzmann statistics, we can write the probability of a spin being in state $\pm 1$ as
$P(\pm1) \sim \exp{(\pm\beta(h_i + J\sigma_{tr}))}$.
Assuming a smooth and continuous profile for $s = \langle \sigma \rangle \approx \langle \sigma_{tr} \rangle$, and the definition, $s = P(+1)-P(-1)$, we obtain by self consistency $s = \tanh{(\beta(hx+2Js))}$.
Near the interface, $x=0, s=0$, the slope is given by,
\begin{equation}
    \frac{ds}{dx} = \frac{\beta h}{1- 2\beta J}.
\end{equation}
The slope diverges for $J=J_*=1/2\beta$, the optimal interaction strength for which the system depicts maximal precision [Fig.~\ref{fig4}~(c)]. Near the interface, the spatial profile is [as indicated
in Fig.~\ref{fig4}~(b)]:
\begin{equation}
    s_* = \frac{1}{2J}\left(\frac{z_*}{\beta}-hx\right),
\end{equation}
where
\begin{equation}
\begin{split}
    z_* = \left(\frac{3hx}{4J} + \left(\frac{9h^2x^2}{16J^2} + \frac{(1-2\beta J)^3}{8\beta^3 J^3}\right)\right)^{1/3} \\ + \left(\frac{3hx}{4J} - \left(\frac{9h^2x^2}{16J^2} + \frac{(1-2\beta J)^3}{8\beta^3 J^3}\right)\right)^{1/3}.
\end{split}
\end{equation}
It should be noted that increasing the temperature $T$ associated with the free energy of individual cells improves the robustness provided $T\geq T_c$, but at the coarse-grained tissue scale, robustness is observed when $T \leq 2J$, implying these two ``effective temperatures" are distinct quantities and are inversely related.
\section{Discussion}
We demonstrate here the appropriateness of invoking free energy landscapes (that
have earlier been seen as analogous to the Waddington landscape) to describe 
cell fate determination. The equilibrium description makes the system highly tractable, while the non-equilibirum aspects of cellular dynamics are incorporated as explicit state-dependent mechanisms that remodels the landscape. 
Symmetry breaking as a mechanism of cell-fate determination is prone to errors in the 
developmental trajectory that arise from fluctuations in gene expression and environmental signals. Proper functionality requires well orchestrated developmental processes, which further demands robust decision making in cells. When confronted with noise, precision requires either integrating signals in space or time. Barring the differences in energetic cost, spatial integration is often more advantageous, because natural selection not only screens for organisms with precise developmental trajectories, but also individuals who mature and reproduce faster. Control over the patterning time can itself be a source of phenotypic variability in a developing embryo without necessarily having to rely on novelty via mutations~\cite{de1951embryos, 10.1007/978-3-642-45532-2_16}. If we consider only the tissue segmentation process, variability in domain proportions come at the expense of precise localization of the interface. However, it is essential to constantly adapt to a dynamic fitness landscape~\cite{West-Eberhard1989}. In the presence of cell-cell communication, we demonstrate minimal non-zero variability in domain proportions with sharp interfaces nevertheless. Taking it a step further, selection pressure also selects for evolvability.
Contact-mediated interactions are capable of assimilating independent mechanisms that can tune the epigenetic landscape to criticality, even if it gets repeatedly perturbed by genetic mutations or environmental variability. In cases where $T\geq T_c$, juxtacrine signaling allows viability and the 
capacity to reproduce. Assuming that the landscape corresponds to a heritable trait ($T$ is inherited from parent to offspring, with small gradual variations), selection for robust localization and interface sharpness will tune the system to criticality $T\rightarrow T_c$. An asymptotic outcome of such selection pressure is to select for regulatory schemes that enables self-organized criticality, where
sufficient genetic buffer canalizes over a large variety of genotypic variations and environmental fluctuations, retaining the fine-tuned landscapes in undifferentiated cells~\cite{doi:10.1073/pnas.0404656101, doi:10.1073/pnas.2413930121, Volkening2018, WADDINGTON1942}.

Waddington's metaphor allows us to rethink the structure-function problem in biology. For instance, the downstream targets in Eph-Ephrin pathway structurally diverges for the bi-directional response, while
they tend to be convergent in terms of functionality~\cite{DRESCHER2002397}. The actual regulatory mechanisms might be more complicated than the one we have assumed here, but the properties of the landscape that emerges out of these descriptions are indifferent to such details. In fact, there could be various mechanisms in the cell, all of which work cooperatively towards multiple goals in a distributed fashion (such as regulating the cell cycle, modulating migration in the extra-cellular matrix, controlling proliferation), and yet they can be successfully integrated in the energy landscape description. The interface dynamics that captures the effect of trans-signaling can be equally valid as a description for cell sorting, a well-known mechanism for segmentation. Eph-Ephrin pathway is known to regulate adhesion and contractility in cells by modulating cell migration~\cite{Zimmer2003, Weijer2009, Amack2012, Hiraiwa2020}. Assuming a minimal dynamical model in which similar cell types 
prefer to stay in proximity of each other, this would be analogous to including an interaction term $J_{ij} = J(1+S_iS_j)$, which reduces (increases) the energy of homogeneous (jagged) patterns. Despite being different from the interaction term we had considered in the case of cell-signaling, viz., $J_{ij} = J(1-S_iS_j)$, the free energy description that models the epigenetic landscape will remain the same. Thus, very distinct mechanisms, even if they operate independently, could be unified in this framework, provided they contribute to similar phenotypic outcomes. This is possible because landscape representations at the cellular scale that may have divergent responses for different regulatory schemes will nevertheless converge to the same coarse-grained description at the tissue scale. 

\begin{acknowledgments}
We would like to thank Chandrashekar Kuyyamudi, Shakti N. Menon, Archishman Raju and
Mohit Kumar Jolly for 
helpful discussions. 
This research has been supported in part by the Center of Excellence in Complex
Systems and Data Science, funded by the Department of Atomic Energy, Government of India.
\end{acknowledgments}
\bibliography{references}

\end{document}